\documentclass[prb,showpacs,twocolumn,preprintnumbers,amsmath,amssymb,floatfix]{revtex4}
\usepackage[dvips]{graphicx}
\usepackage[latin1]{inputenc}
\begin{document}
\draft

\newcommand{\lxpc} {Li$_{x}$ZnPc }
\newcommand{\lpc} {Li$_{0.5}$MnPc }
\newcommand{\etal} {{\it et al.} }
\newcommand{\ie} {{\it i.e.} }
\newcommand{\ip}{${\cal A}^2$ }

\hyphenation{a-long}

\title{NMR and $\mu$SR study of spin correlations in SrZnVO(PO$_4$)$_2$ a S=1/2 frustrated magnet on a square lattice }

\author{L. Bossoni$^{1}$, P. Carretta$^{1}$, R. Nath$^{2}$, M. Moscardini$^{1}$, M. Baenitz$^{3}$  and C. Geibel$^{3}$.}

\address{$^{1}$ Department of Physics ``A. Volta'', University of Pavia-CNISM, 27100 Pavia (Italy)}
\address{$^{2}$ Indian Institute of Science Education and Research-Thiruvananthapuram, 695016 Kerala (India)}
\address{$^{3}$ Max-Planck Institute for Chemical Physics of Solids, 01187 Dresden (Germany)}

\widetext

\begin{abstract}

$^{31}$P nuclear and muon spin-lattice relaxation rate measurements in SrZnVO(PO$_4$)$_2$, a $S=1/2$ frustrated
magnet on a square lattice, are presented. The temperature ($T$) dependence of the in-plane correlation length
$\xi$ is derived and it is shown that the overall behaviour is analogous to the one found for non-frustrated
systems but with a reduced spin stiffness. The temperature dependence of $\xi$ in SrZnVO(PO$_4$)$_2$ is compared
to the one of other frustrated magnets on a square lattice with competing nearest neighbour ($J_1$) and
next-nearest neighbour ($J_2$) exchange couplings and it is shown that $\xi$ progressively decreases as the ratio
$J_2/J_1$ approaches the critical value leading to the suppression of long-range magnetic order. In spite of the
differences in the functional form of $\xi(T)$ found in different vanadates, it is pointed out that the
characteristic energy scale describing spin correlations in all those compounds appears to scale as $|2J_2 +
J_1|$.

\end{abstract}

 \maketitle

\narrowtext

\section{Introduction}
The study of quantum magnetism has received a renewed attention after the discovery of high temperature
superconductivity in the cuprates. In fact, these materials have allowed to investigate at the experimental level
the phase diagram of S=1/2 Heisenberg antiferromagnets on a square lattice with great accuracy \cite{Cuprates}.
The behaviour of the correlation length has been derived by means of neutron scattering experiments \cite{Keimer}
and nuclear spin-lattice relaxation rate measurements \cite{Imai,PC1}, the form of the dynamical susceptibility
and, accordingly, the value of the scaling exponents which characterize those systems have been obtained.
\cite{Birgenau} More recently much attention has been addressed to the investigation of frustrated square lattice
(FSL) systems where the frustration is induced by a next nearest neighbour (n.n.n.) exchange coupling $J_2$
competing with the nearest neighbour (n.n.) one ($J_1$) along the side of the square \cite{J1J2}. Frustration is
expected to further enhance quantum fluctuations and to lead to the suppression of long-range magnetic order for
certain values of the ratio $r=J_2/J_1$. In particular, when both exchange couplings are antiferromagnetic for
$r\simeq 0.5$ a spin-liquid ground-state is expected \cite{TeoJ1J2}, while when $J_1$ is ferromagnetic (i.e. $J_1<
0$) for $r\simeq -0.5$ a nematic order for the two-spin correlation function is envisaged.\cite{Nic} The $J_1-J_2$
model on a square lattice has received renewed attention in the last two years when it was realized that the
parent compounds of the recently discovered iron-based superconductors are characterized by comparable n.n. and
n.n.n. hopping integrals, which may yield competing exchange couplings within the square lattice formed by iron
atoms\cite{Pnict}. In fact, those materials would represent an extension of the $J_1-J_2$ model on a square
lattice to itinerant electron systems. As regards the insulating systems, a number of compounds have been recently
identified to be prototypes of FSL systems and investigated through different experimental
approaches.\cite{J1J2,TeoRec,Synthesis,BaCd,Tsi1,Tsi2,Skou} Attempts have been made to theoretically understand
their high field properties \cite{TeoRec,Thalm} and the exchange mechanisms.\cite{Tsirlin} Despite such a
theoretical and experimental progress, a number of key questions still have to be addressed. For example, the
parts of the phase diagram where long-range order should be absent have not been studied so far, moreover it is
not clear how the temperature dependence of the in-plane correlation length $\xi$ changes with $r$. Unfortunately,
it is not possible to address this latter aspect by means of inelastic neutron scattering experiments since only
small crystals are available for the prototypes of the $J_1-J_2$ model on a square lattice \cite{J1J2}. Hence, it
would be worthwhile to find other experimental techniques which could allow to determine the effect of frustration
on $\xi$.

Here we present an experimental study of the temperature ($T$) dependence of the in-plane correlation length
$\xi$, derived by means of nuclear and muon spin-lattice relaxation rates, in SrZnVO(PO$_4$)$_2$
(Fig.\ref{Strutt})  a prototype of frustrated magnet on a square lattice with competing ferromagnetic n.n. and
antiferromagnetic n.n.n. couplings. It will be shown that $\xi$ diverges exponentially on cooling with a reduced
spin stiffness, possibly scaling as $|J_1+ 2J_2|$. A comparison with the results previously obtained by our group
on other systems with $r< 0$ appears to qualitatively support this scaling of the spin stiffness, even if an
accurate description of $\xi$ on approaching the transition to the columnar ground-state should take into account
the spin anisotropy and interlayer couplings.

\section{Technical aspects and Experimental Results}

The synthesis of SrZnVO(PO$_4$)$_2$ polycrystalline sample was carried out by using the protocol already reported
in Refs.\onlinecite{Synthesis} and \onlinecite{Structure}. DC magnetization ($M$) measurements were performed in
order to estimate the superexchange coupling constants for our sample and to check if they are consistent with the
ones reported in the literature.\cite{TeoRec,Synthesis,Tsirlin} The $T$-dependence of the static uniform spin
susceptibility $\chi= M/H$,\cite{Suppl} with $H$ the magnetic field intensity, was analyzed by fitting the
high-$T$ data to Curie-Weiss law and to the high-$T$ series expansion.\cite{HTE} It was found that, $J_1= -7.53
\pm 0.7$ K and $J_2= 8.63 \pm 0.6$ K, values which are quite consistent with the ones previously reported in the
literature.\cite{TeoRec,Synthesis,Tsirlin} In the following, in order to better compare SrZnVO(PO$_4$)$_2$ to the
other systems we shall introduce a characteristic energy scale $J_C=\sqrt{J_1^2+J_2^2}\simeq 11.45$ K, which
provides the magnitude of the exchange couplings.

\begin{figure}[h!]
\vspace{6cm} \includegraphics{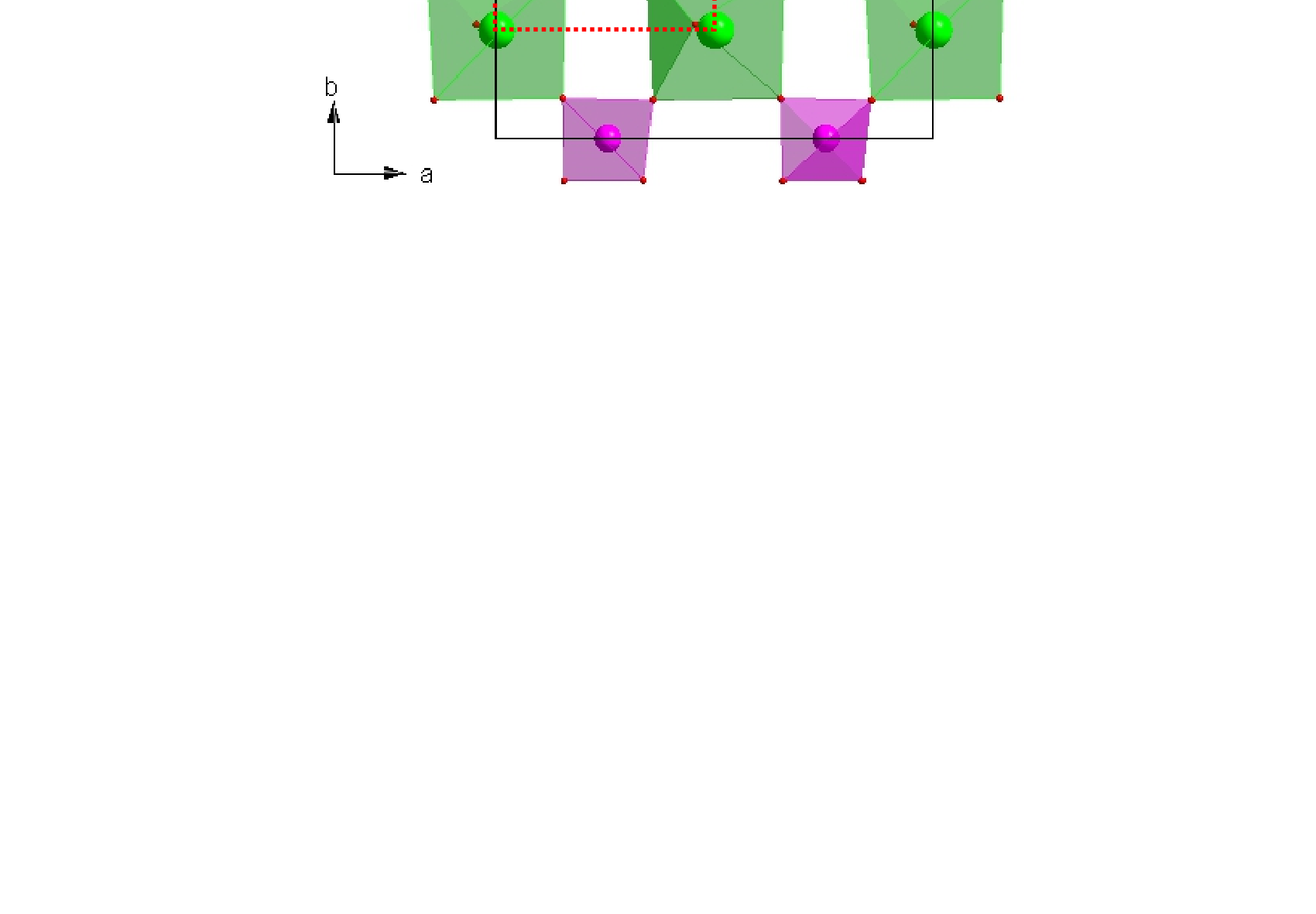} \caption{\label{Strutt}
Projection of SrZnVO(PO$_4$)$_2$ structure along the c axis evidencing the planes containing V$^{4+}$ ions.
Vanadium ions are in green, phosphorus in purple and oxygen ions in red. VO$_5$ pyramids and P1O$_4$ tetrahedra
are also visible. The red dotted square shows the $S=1/2$ square lattice. }
\end{figure}

\begin{figure}[h!]
\vspace{13cm} \includegraphics{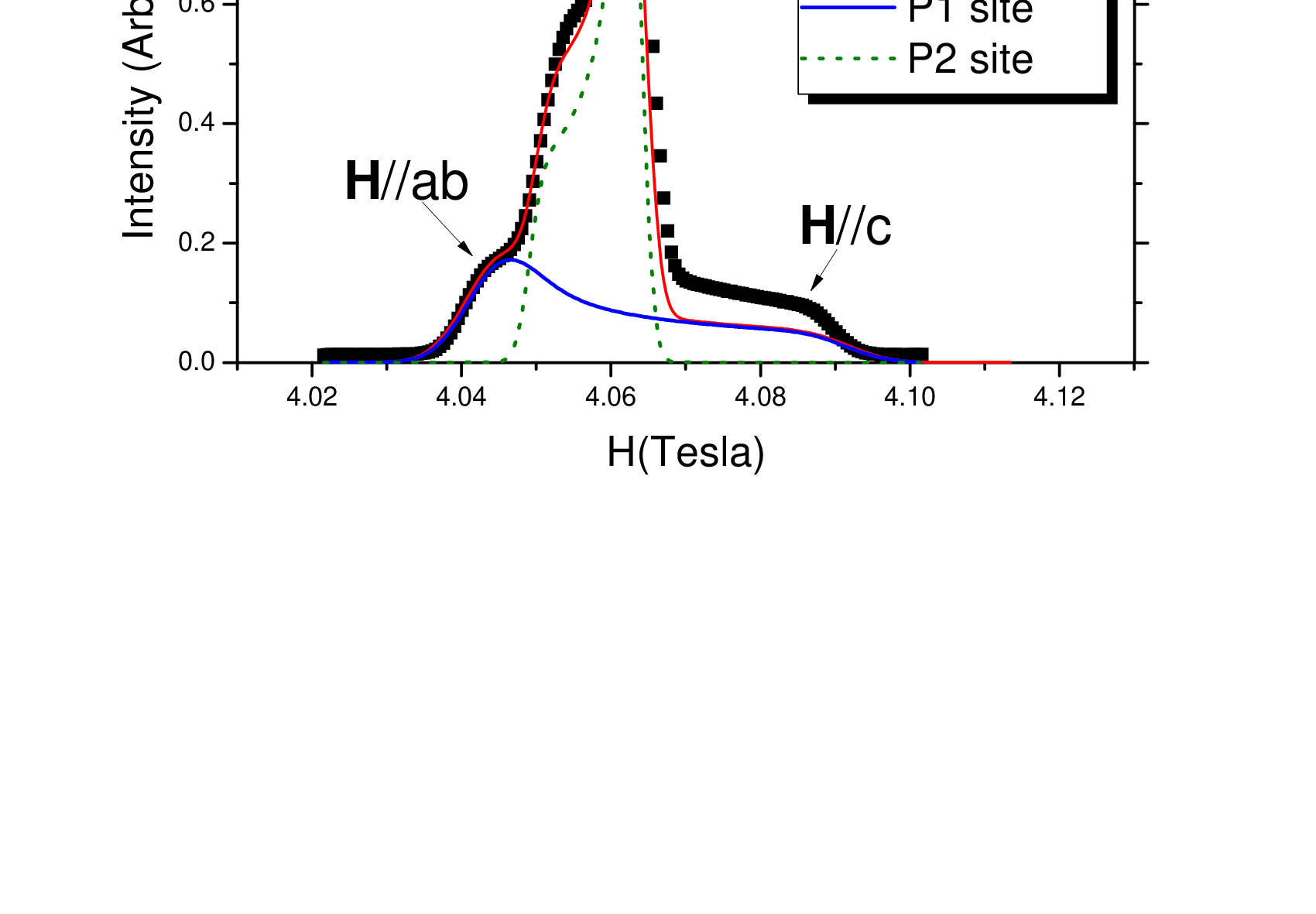} \includegraphics{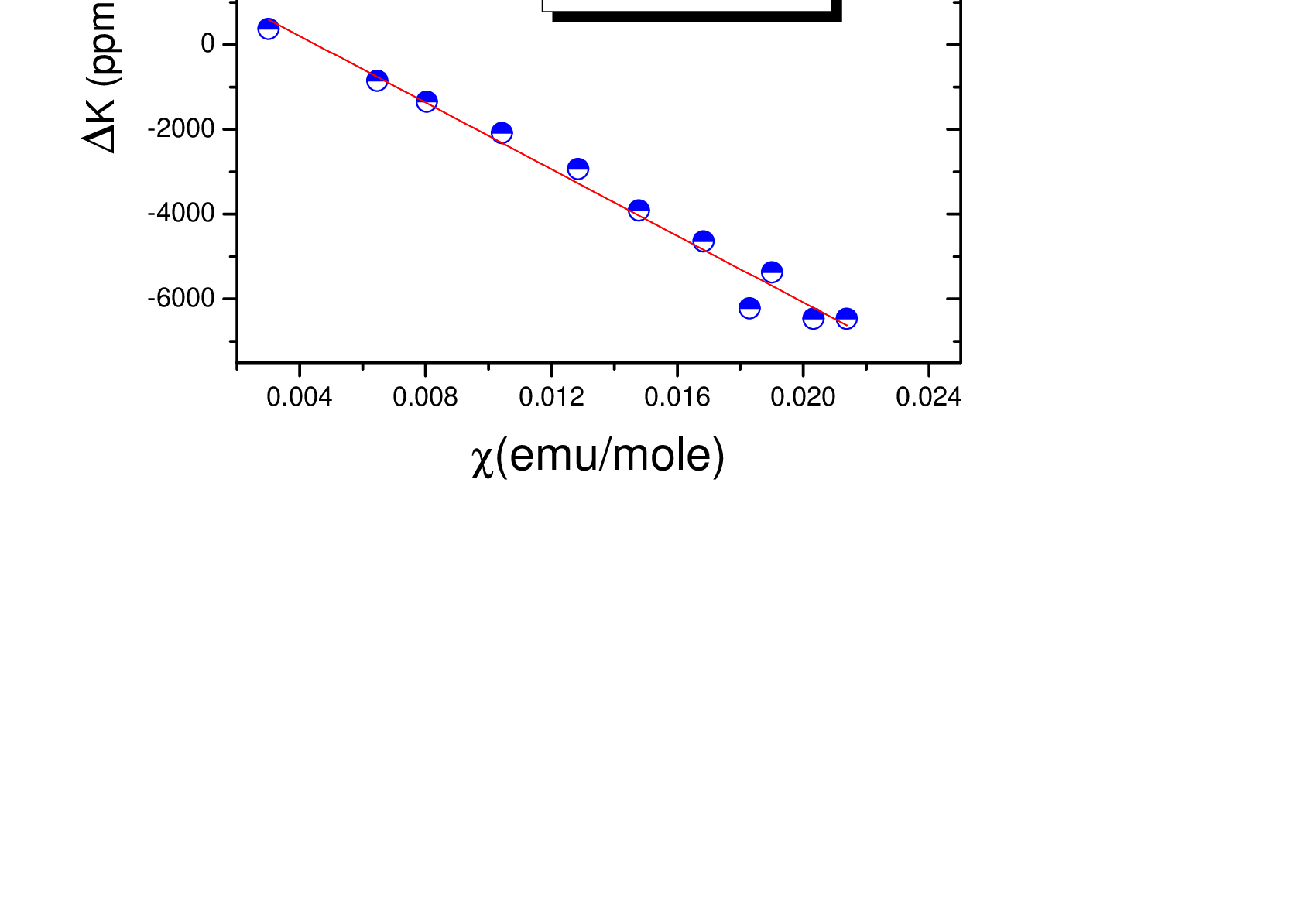} \caption{\label{shift} (a) Field swept $^{31}$P NMR spectrum is
reported for RF irradiation at $\nu= 70$ MHz. The contribution from $^{31}$P1 and $^{31}$P2 sites is evidenced and
the parts of $^{31}$P1 spectra corresponding to an orientation of the grains with $\mathbf{H}\parallel c$ or
$\mathbf{H}\parallel ab$ are shown. (b) The $^{31}$P NMR shift of the high (green) and low (blue) frequency
shoulders of $^{31}$P1 NMR spectra, corresponding to $\mathbf{H}\parallel ab$ and $\mathbf{H}\parallel c$,
respectively, is reported as a function of the macroscopic spin susceptibility with the temperature as an implicit
parameter. }
\end{figure}

$^{31}$P NMR measurements were carried out by using standard radiofrequency (RF) pulse sequences. At low field,
where the full spectrum could be irradiated, the NMR powder spectra were obtained from the Fourier transform of
half of the echo after a $\pi/2-\tau_E-\pi$ pulse sequence. The NMR powder spectrum Fig. \ref{shift}(a) was
characterized by an asymmetric line shape, quite similar to the one found by Nath et al.\cite{Nath} in the
isostructural Pb$_2$VO(PO$_4$)$_2$ compound. The narrow central component is associated with P2 sites lying in
between adjacent vanadium layers, while the broader component to the P1 site (Fig. \ref{Strutt}) which lies within
vanadium layers and is characterized by a larger hyperfine coupling. At high magnetic fields the line broadening
prevented the irradiation of the whole line and the NMR spectrum had to be derived either by recording the
intensity of the signal upon making discrete frequency steps or upon sweeping the magnetic field (Fig.
\ref{shift}(a)). The T-dependence of the NMR shift $\Delta K$ for the P1 site for $\mathbf{H}\parallel c$ and
$\mathbf{H}\parallel ab$ was determined by recording the position of the low frequency (high field) and of the
high frequency (low field) shoulders of the NMR spectrum, respectively, as a function of T. Both quantities were
found to scale linearly with $\chi$ but with opposite slopes, indicating an opposite sign in the hyperfine
coupling components (Fig. \ref{shift}(b)).

\begin{figure}[h!]
\vspace{7cm} \includegraphics{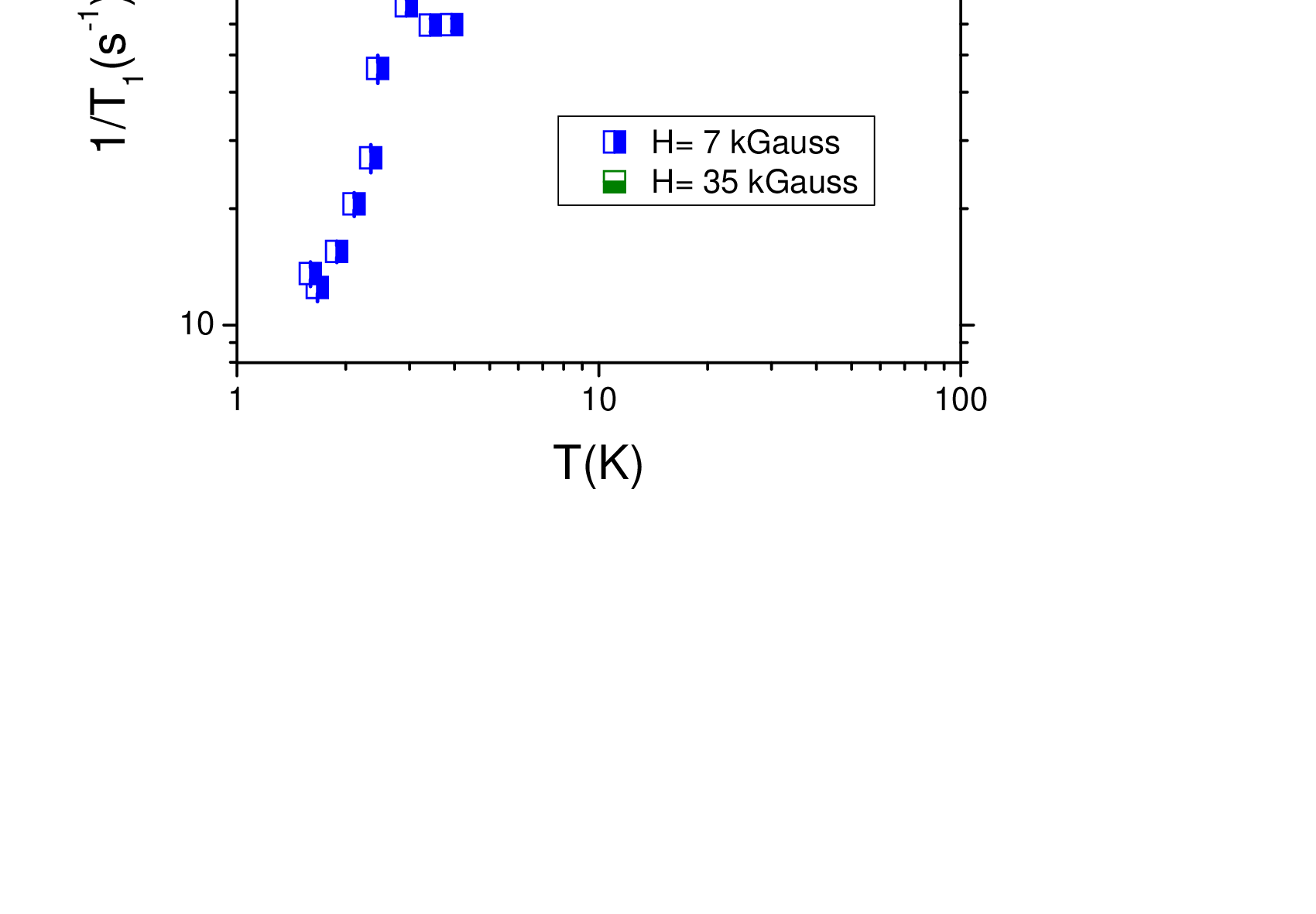} \caption{\label{T1vsT} Temperature
dependence of $^{31}$P1 nuclear spin-lattice relaxation rate in
SrZnVO(PO$_4$)$_2$.}
\end{figure}

Nuclear spin-lattice relaxation rate $1/T_1$ was derived from the recovery of the nuclear magnetization after a
saturating pulse sequence. In view of the anisotropy of the hyperfine coupling tensor $1/T_1$ depends on the
portion of the spectrum being irradiated. Since the low-frequency (high field) shoulder of the $^{31}$P1 spectrum
was more separated from the rest of the spectra we have decided to irradiate just that part of $^{31}$P NMR powder
spectrum, corresponding to the crystallites with $\mathbf{H}\parallel c$. From now on we shall refer to $T_1$ only
for the $^{31}$P1 site and for that orientation. The corresponding recovery laws for the nuclear magnetization
could be nicely fit by a single exponential. In Fig. \ref{T1vsT} the temperature dependence of $1/T_1$ in the 1.6
K - 100 K range is shown. At high temperature $1/T_1$ is flat, then it smoothly decreases and eventually it shows
a well defined peak at $T_C= 2.65\pm 0.02$ K, corresponding to the columnar ordering temperature. Below $T_C$ a
rapid decrease of $1/T_1$ is observed. This behaviour is very similar to the one reported by Nath et al.
\cite{Nath} for Pb$_2$VO(PO$_4$)$_2$. No significant change in $1/T_1$ was noticed upon increasing the magnetic
field intensity from 7 to 35 kGauss (Fig.\ref{T1vsT}) at high $T$. On the other hand, a tiny change has to be
expected for $T\rightarrow T_C$ due to the variation of the transition temperature with the field.\cite{Nath}

\begin{figure}[h!]
\vspace{7cm} \includegraphics{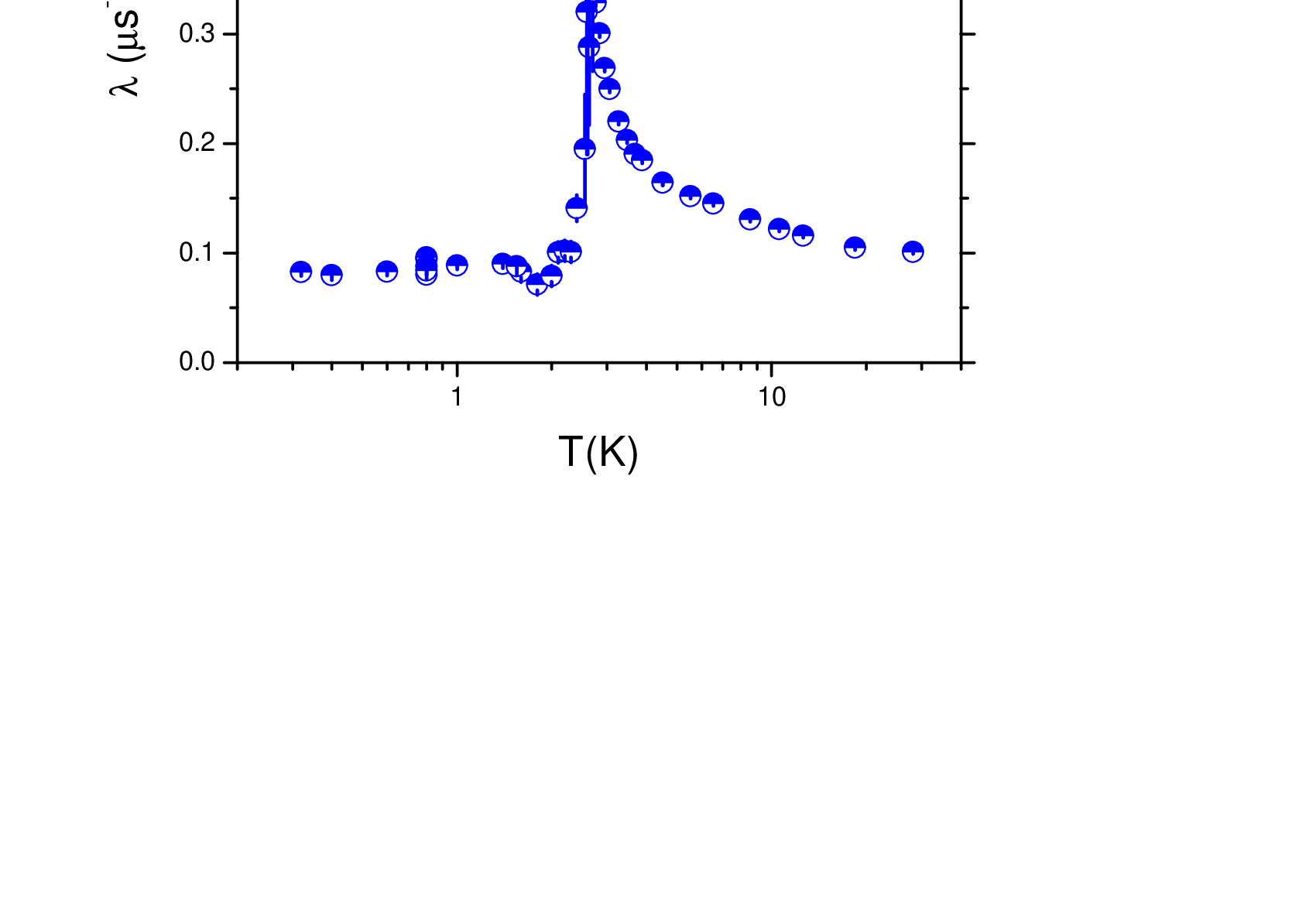} \caption{\label{lambda}
Temperature dependence of the zero-field muon relaxation rate in SrZnVO(PO$_4$)$_2$.}
\end{figure}

$\mu$SR measurements were performed at ISIS pulsed muon source on MUSR beam line. In zero-field (ZF), for $T>
T_C$, the decay of the muon asymmetry was characterized by a stretched exponential function $A(t)=
A(0)exp(-(\lambda t)^{\beta})$,\cite{Suppl} with $\beta$ progressively decreasing from $0.7$ to $0.5$ upon
decreasing the temperature from 30 K to $T_C$. The stretched exponential character of the relaxation can be
associated either with a distribution of muon sites or with an anisotropic hyperfine coupling, yielding to a
distribution of relaxation rates in a powder sample. Below $T_C$ clear oscillations are observed in
zero-field,\cite{Suppl} showing that there is a spontaneous sublattice magnetization causing a non-zero magnetic
field $B_{\mu}$ at the muon site \cite{PC2}. Accordingly the decay of the muon asymmetry followed the behaviour
typically found for powder samples in zero-field \cite{Blundell}

\begin{equation}
\label{ZFAsymm}
      A(t)= A_1 e^{-\sigma t} cos(\gamma_{\mu}B_{\mu}t + \phi) +
      A_2 e^{-\lambda t} +B
  \;\;\;,
\end{equation}
where $\gamma_{\mu}$ is the muon gyromagnetic ratio, $\sigma$ the decay rate of the oscillating part, mostly due
to a inhomogeneous distribution of the local field at the muon sites, while $B$ is a constant background arising
from the sample environment.

\begin{figure}[h!]
\vspace{7cm} \includegraphics{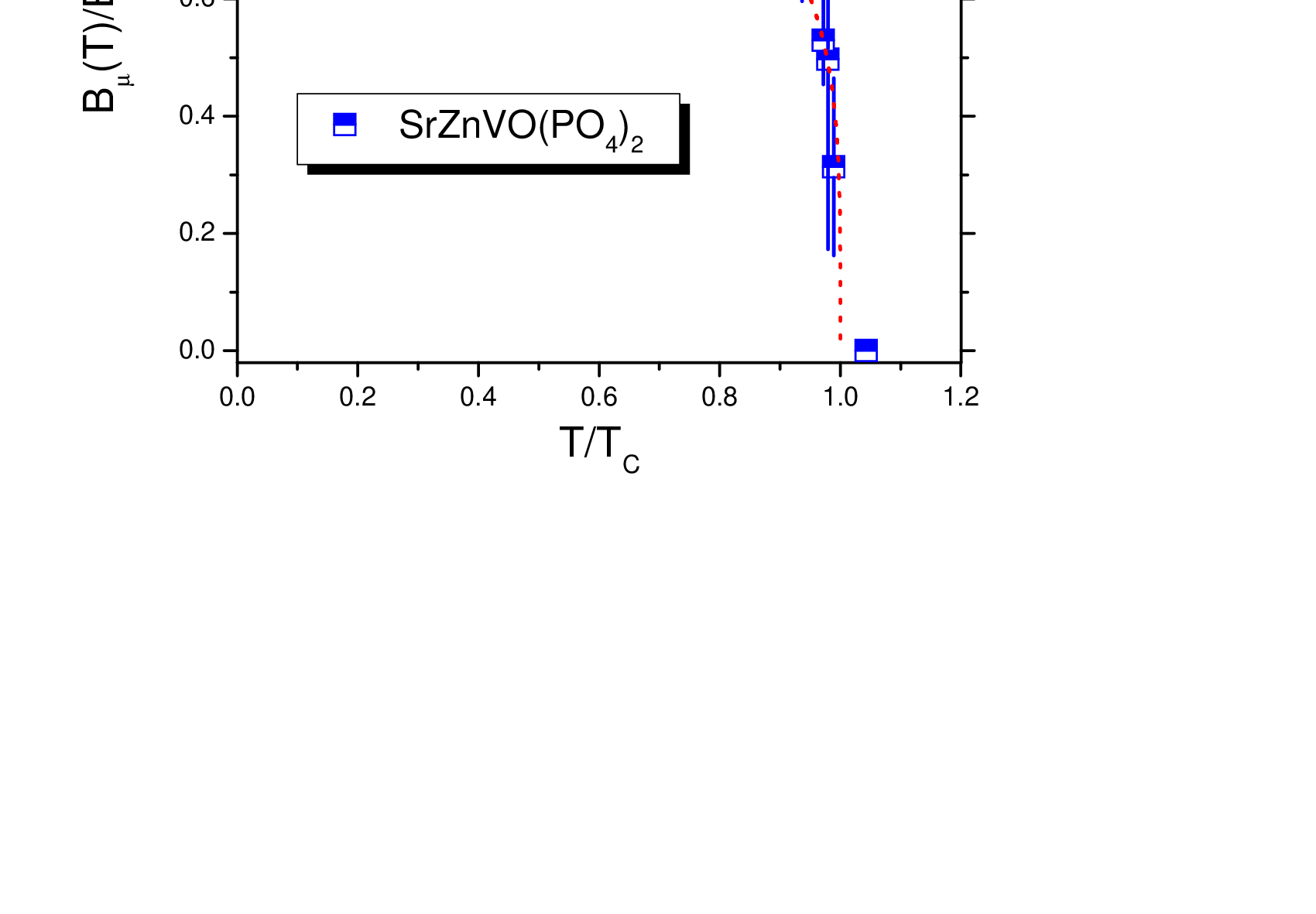} \caption{\label{Bmu} The
local field at the muon, normalized to its low temperature value $B_{\mu}= 186\pm 3$ Gauss, is reported as a
function of $T/T_c$, with $T_c= 2.65$ K. The low-temperature blue dotted line shows the behaviour expected for a
power law reduction of $B_{\mu}(T)=B_{\mu}(0)(1 - a T^2)$. The high temperature red dotted line shows the critical
behaviour for a critical exponent $\beta= 0.235$, expected for a 2D XY model.}
\end{figure}

The temperature dependence of $\lambda$ and of $B_{\mu}$ derived from the fit of the asymmetry with the
aforementioned expressions is reported in Fig. \ref{lambda} and Fig. \ref{Bmu}, respectively.

\section{Discussion}

First we shall consider the temperature dependence of the order parameter, as derived from zero-field $\mu$SR
measurements. The local field at the muon can be written
\begin{equation}
B_{\mu}= \sum_i A^{\mu}_i<\vec S_i>= A^{\mu}_{eff} |<\vec S>|
  \;\;\;,
\end{equation}
where $A^{\mu}_i$ is the hyperfine coupling between the muon and the i-th V$^{4+}$ spin. Since the magnitude of
all V$^{4+}$ spins $|<\vec S>|$ is expected to be the same, the local field at the muon can be written in terms of
an effective total hyperfine coupling $A^{\mu}_{eff}$ times $|<\vec S>|$. Hence, the $T$-dependence of the local
field at the muon gives directly the one of the sublattice magnetization. As regards, the critical behaviour of
the order parameter, here we only remark that the critical exponent $\beta$ can be consistent with the one
expected for finite 2DXY systems ($\beta= 0.235$)\cite{Bram} found in other similar vanadates.\cite{Nath,PC2,PC3}
Nevertheless, the accuracy of the experimental points does not allow one to give a definite answer in this
respect. On the other hand, the low temperature behaviour of $B_{\mu}$ provides information on the dispersion
relation for the spin waves excitations. The reduction of the low-temperature sublattice magnetization is
consistent with a power-law $|<\vec S>|(T)\sim T^n$, with $n=1.8\pm 0.4$. Although this value would be consistent
with the dispersion relation for nearly two-dimensional antiferromagnets,\cite{Bucci}  it is difficult to give a
precise statement in view of the experimental uncertainty. Nevertheless, as it will be shown at the end of this
section, also the $T$-dependence of $1/T_1$ seems to support the 2D character of the spin wave excitations, with a
quasi-linear magnon dispersion.

Now we turn to the discussion of the temperature dependence of $^{31}$P nuclear spin-lattice relaxation rate,
which allows to derive information on the low-energy dynamics and on the spin correlations. In the case of a
magnetic relaxation process driven by electron spin fluctuations $1/T_1$ can be written \cite{Moriya}
\begin{equation}
\frac{1}{T_1}= \frac{\gamma^2}{2N}\sum_{\alpha,\mathbf{q}} (|A_{\mathbf{q}}|^2
S_{\alpha,\alpha}(\mathbf{q},\omega_L))_{\perp}
  \;\;\;,
\end{equation}
where $\gamma$ is the nuclear gyromagnetic ratio, $|A_{\mathbf{q}}|^2$ the form factor describing the hyperfine
coupling with spin excitations at wave-vector $\mathbf{q}$ and $S_{\alpha,\alpha}(\mathbf{q},\omega_L)$
($\alpha=x,y,z$) the component of the dynamical structure factor at the Larmor frequency $\omega_L$. The $\perp$
subscript indicates that one should consider the components of the fluctuating hyperfine field perpendicular to
the quantization axis, given by the direction of the static external field. By using scaling arguments it is
possible to write the dynamical structure factor in terms of the in-plane correlation length $\xi$ (in lattice
units hereafter) and establish a one to one relationship between $1/T_1$ and $\xi$. This procedure has proven to
be very useful to study the temperature dependence of the correlation length in the cuprates which are prototypes
of two-dimensional $S=1/2$ Heisenberg antiferromagnets on a square lattice \cite{PC1} and to determine the value
of the dynamical scaling exponent $z=1$. Given the similarity between the cuprates and the vanadates under
investigation, we will use the same approach here to derive $\xi$ from $^{31}$P $1/T_1$, assuming $z=1$.
Accordingly, one can write \cite{PC1}
\begin{equation}
\label{scale} \frac{1}{T_1}\simeq {\gamma^2} \frac{S(S+1)}{3} \xi^{z+2} \frac{\beta(\xi)^2\sqrt{2\pi}}{\omega_E}
\frac{1}{4\pi^2}\times \nonumber
\end{equation}
\begin{equation}
\times\int_{BZ} d\vec q \frac{|A_{\vec q}|^2 }{1 + \xi^2(\vec q - \vec Q_C)^2}
  \;\;\;,
\end{equation}
where $\beta$ is a normalization factor which allows to preserve the spin sum rule and $\vec Q_C$ is the columnar
critical wave-vector. $\omega_E= (J_Ck_B/\hbar)\sqrt{2nS(S+1)/3}$ is the Heisenberg exchange
frequency,\cite{Moriya} with $n=4$ the number of n.n. and of next n.n..

In order to establish a one to one relationship between $1/T_1$ and $\xi$ from the above equation one has first to
derive the hyperfine coupling tensor components and the form factor for $^{31}$P1 site. The components of the
hyperfine tensor can be derived from the plot of the shift $\Delta K$ for the magnetic field along the $\alpha$
direction vs. the molar macroscopic susceptibility $\chi$ (Fig. 2(b)), which is assumed to be isotropic. Then one
can write
\begin{equation}
\Delta K_{\alpha}= \frac{A_{\alpha\alpha} \chi}{g\mu_B N_A}
  \;\;\,
\end{equation}
with $\mu_B$ the Bohr magneton and $N_A$ the Avogardro's number.

\begin{figure}[h!]
\vspace{6cm} \includegraphics{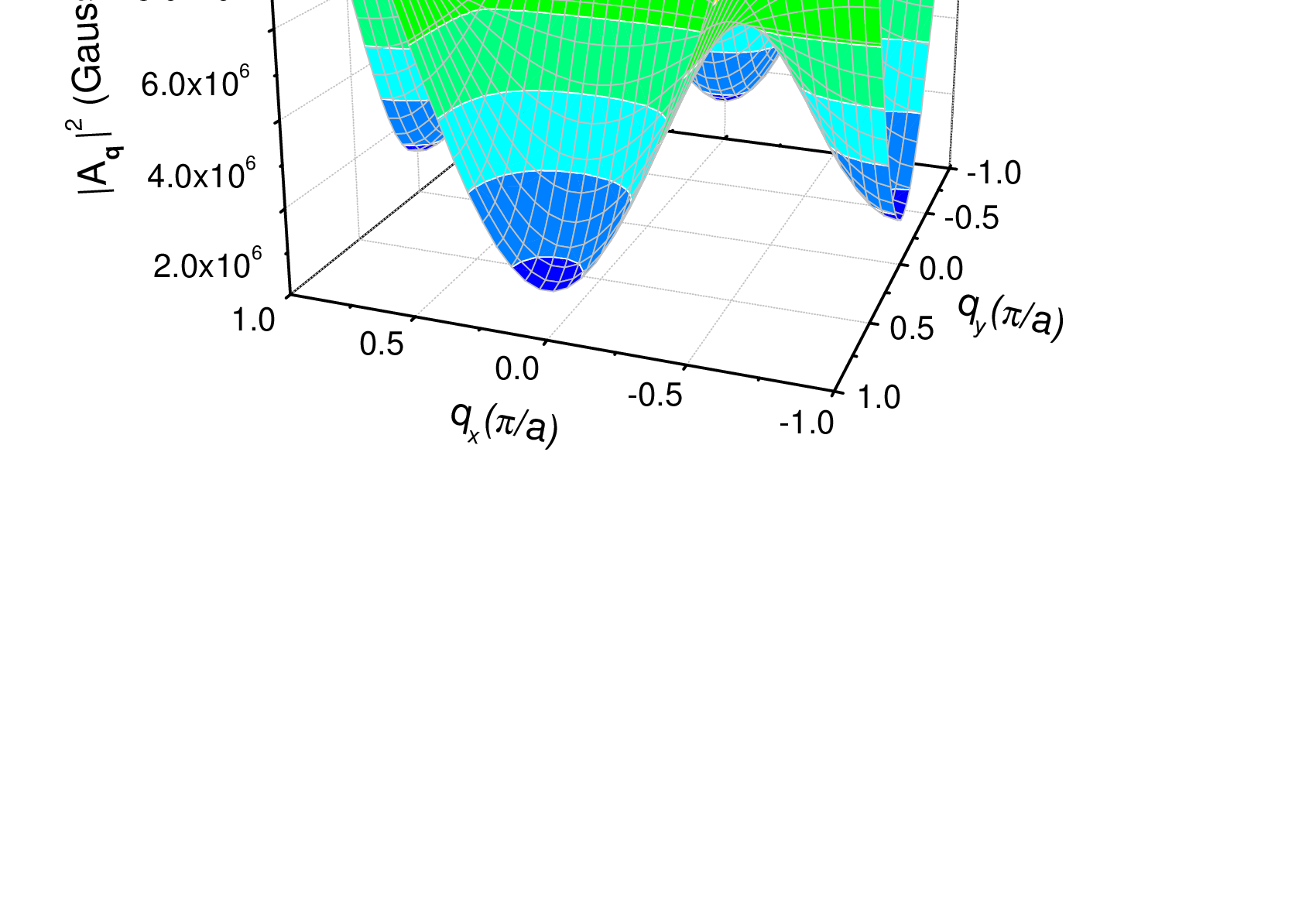} \caption{\label{Form} The form
factors for $^{31}$P1 site are reported as a function of the in-plane components ($q_x$ and $q_y$) of the
wave-vector of the spin excitations for SrZnVO(PO$_4$)$_2$. }
\end{figure}

\begin{figure}[h!]
\vspace{6cm} \includegraphics{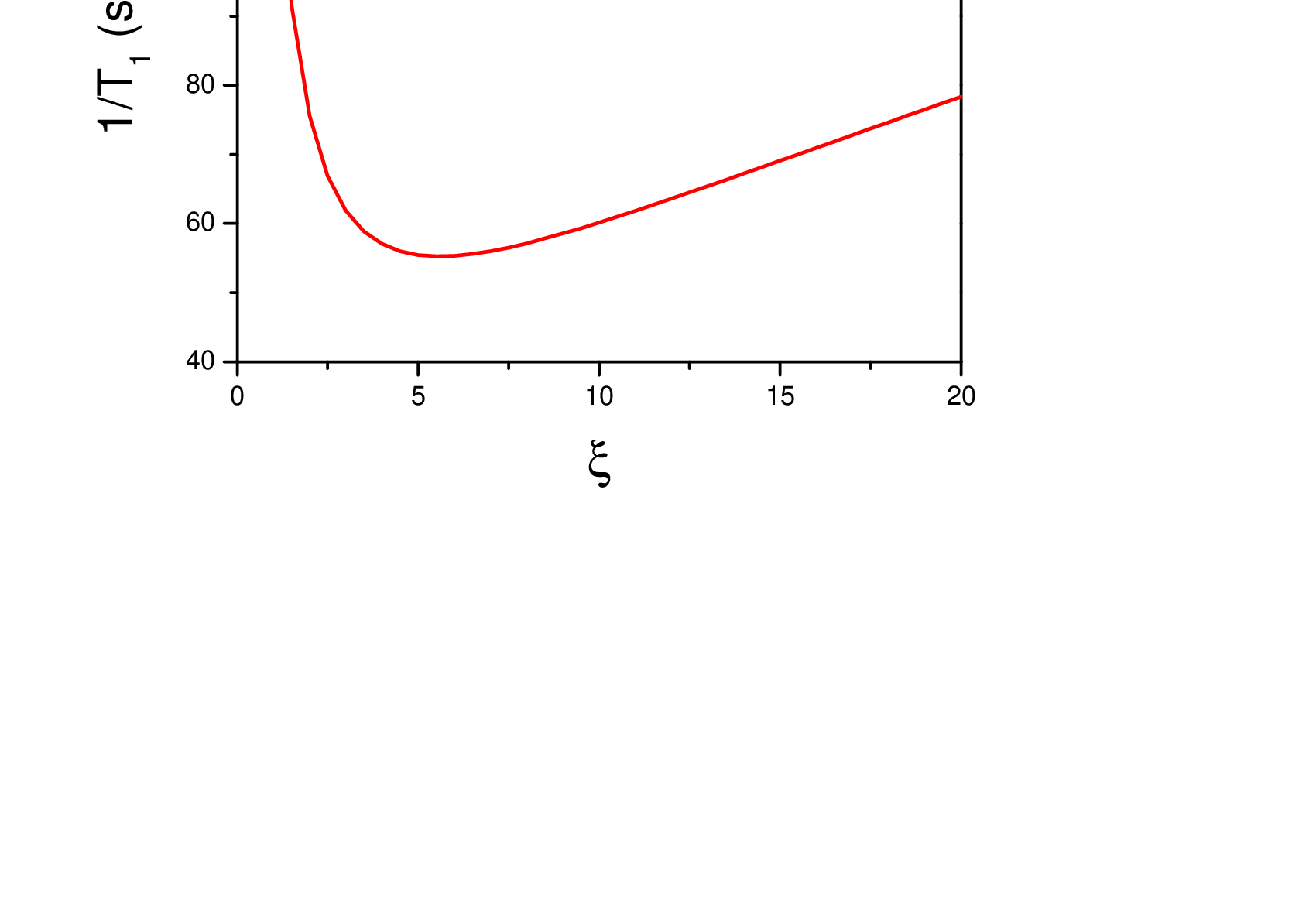} \caption{\label{T1vscsi}
$^{31}$P1 nuclear spin-lattice relaxation rate in SrZnVO(PO$_4$)$_2$ is reported as a function of the in-plane
correlation length (in lattice units) according to Eq.\ref{scale} in the text.}
\end{figure}

For SrZnVO(PO$_4$)$_2$ one finds $A_{cc}= - 4300 \pm 190$ Gauss and $A_{aa}\simeq A_{bb}= 2360\pm 200$ Gauss. In
fact, since the measurements on powders did not allow one to discern between $\Delta K_a$ and $\Delta K_b$ we have
assumed $A_{aa}\simeq A_{bb}$. In Pb$_2$VO(PO$_4$)$_2$, on the other hand, a small difference between those two
components is observed.\cite{Nath} The total hyperfine tensor is the sum of a transferred term $A^t$ and of a
dipolar term $A^{d}$. The latter one can be calculated on the basis of lattice sums while the former one is
assumed to be the sum of four equal terms arising from the hyperfine coupling between $^{31}$P1 nuclei and the
four n.n. V$^{4+}$ spins. Hence the contribution to the transferred hyperfine term is simply given by $A^t= (A -
A^d)/4$. Now that both the transferred and dipolar coupling between $^{31}$P1 nucleus and each V$^{4+}$ spin are
known, it is possible to derive the hyperfine form factor. The form factor of SrZnVO(PO$_4$)$_2$ is  reported in
Fig.\ref{Form}. It is noticed that, owing to the symmetry position of P1 site, the form factor shows a
non-vanishing minimum at $\vec Q_C= (\pm\pi,0)$ and $(0, \pm\pi)$. This explains why $1/T_1$ progressively
decreases as the system gets more correlated upon decreasing temperature (Fig.\ref{T1vsT}) and only when the
correlation length is sufficiently large $1/T_1$ increases again. In fact, from Eq.\ref{scale} it is possible to
derive numerically the behaviour of $1/T_1$ vs. $\xi$ (Fig. \ref{T1vscsi}) and one finds a minimum for $\xi\simeq
5$ lattice steps. Accordingly, from the experimental data reported in Fig. \ref{T1vsT}  it is now possible to
derive quantitatively the temperature dependence of $\xi$ in SrZnVO(PO$_4$)$_2$.

\begin{figure}[h!]
\vspace{7cm} \includegraphics{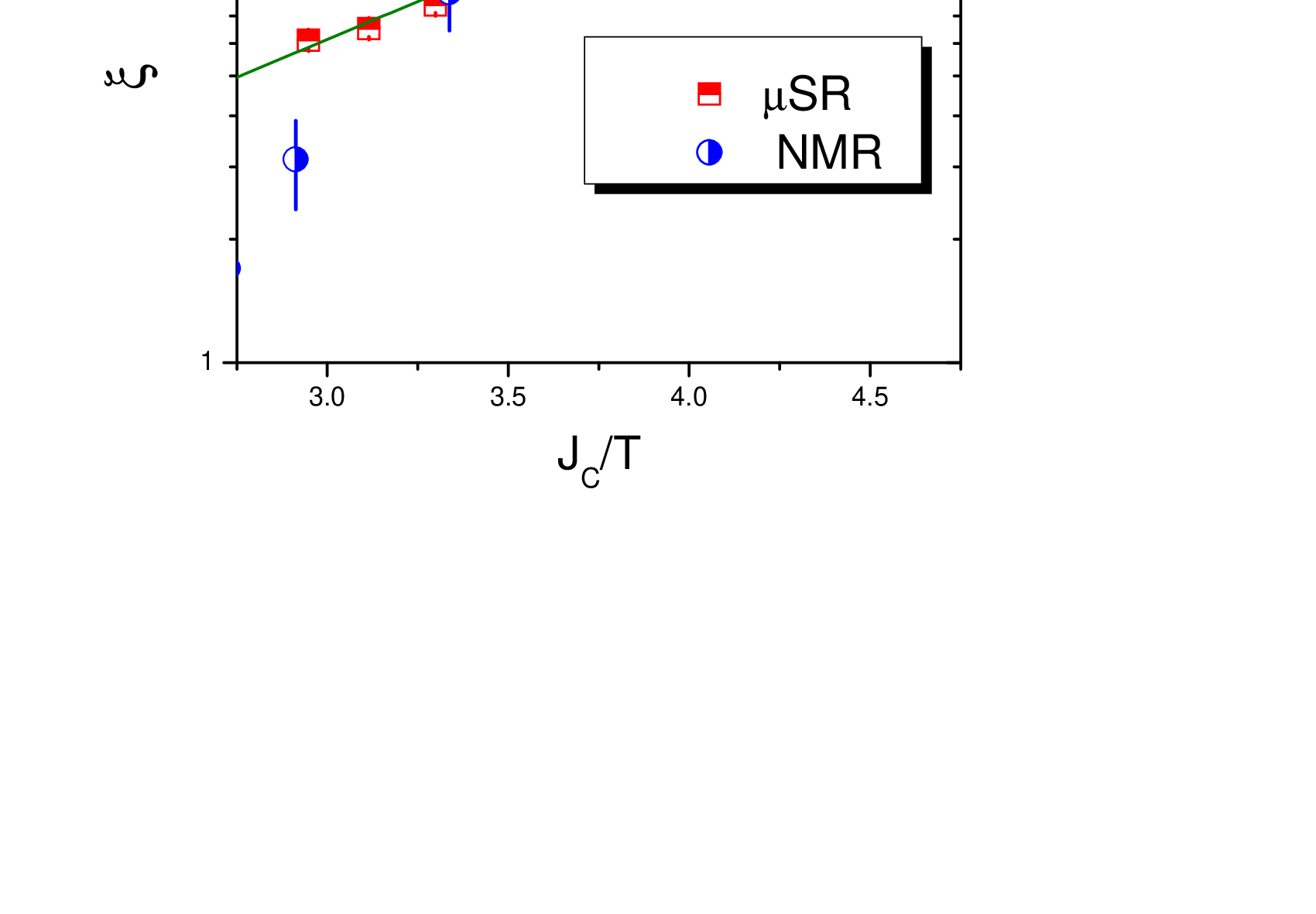} \caption{\label{csivsT} The
temperature dependence of the in-plane correlation length (in lattice units) derived from $\lambda$ and $1/T_1$
data is reported as a function of $J_C/T$. The solid line shows the behaviour expected for a spin-stiffness
$\rho_s= 0.79\pm 0.05\times 1.15 J_C/2\pi$.}
\end{figure}

\begin{figure}[h!]
\vspace{7cm} \includegraphics{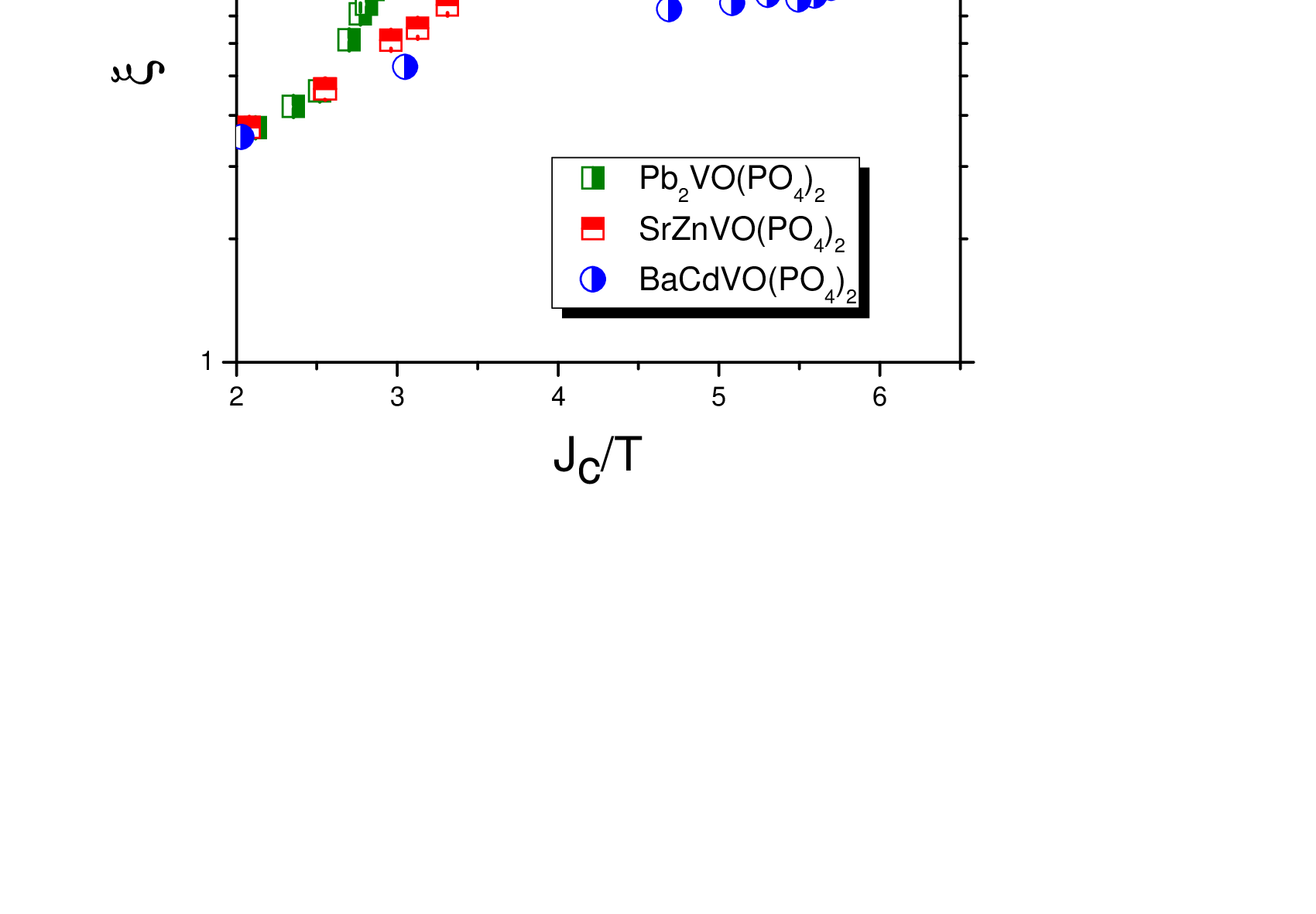} \caption{\label{xiJ2J1} The
temperature dependence  in-plane correlation length derived from $\lambda$ is reported vs. $J_C/T$ for  compared
to the one derived for SrZnVO(PO$_4$)$_2$ ($J_C= 11.45$ K), BaCdVO(PO$_4$)$_2$ ($J_C= 4.8$ K)\cite{BaCd} and
Pb$_2$VO(PO$_4$)$_2$ ($J_C= 10.7$ K)\cite{Pb2}. The dotted arrow points out that upon decreasing $|J_2/J_1|$ the
correlation length increases less rapidly on cooling. }
\end{figure}
In order to derive the temperature dependence of $\xi$ from $\lambda(T)$ one has first to subtract the
$T$-independent dipolar contribution from the raw data in Fig. 4 and then proceed in the same way as it was done
for $1/T_1$. However, here the form factor cannot be determined since the muon site and hyperfine couplings are
unknown. It is noticed that, at variance with P1 site, the muon should not be in a position symmetrical with
respect to the neighbouring V$^{4+}$ ions, since $\lambda$ continues to diverge upon cooling. Thus, if no
filtering effect due to the form factor is present one can safely write $\lambda\sim \xi$, for $\xi\gg
1$.\cite{PC1} Namely, the temperature dependence of $\lambda$ directly gives the one of $\xi$ although, unlike
$^{31}$P $1/T_1$, $\lambda$ does not allow to estimate quantitatively $\xi$. Nevertheless, by matching the
$\xi(T)$ derived from $\lambda(T)$ with the one quantitatively derived from $1/T_1$ over the same $T$ range, it is
possible to use also $\lambda(T)$ data to estimate quantitatively $\xi(T)$.

In Fig.\ref{csivsT} we report the temperature dependence of $\xi$ derived by means of $^{31}$P1 $1/T_1$ and the
one obtained by means of $\lambda(T)$ in a temperature range where $\xi$ is sufficiently large so that either
Eq.\ref{scale} apply or $\lambda\sim \xi$. One notices an overall good agreement in the behaviour of $\xi(T)$
derived through both methods, moreover it is noticed that $\xi$ diverges exponentially on decreasing temperature.
For a non frustrated $S=1/2$ Heisenberg antiferromagnet on a square lattice one would expect that $\xi(T)\simeq
exp(2\pi\rho_s/T)/(T+ 4\pi\rho_s)$,\cite{RCR} with $\rho_s$ the spin stiffness, which for non frustrated systems
turns out $\rho_s\simeq 1.15 J_C/2\pi$. Here we find that the behaviour of $\xi(T)$ is the same but with a reduced
effective spin stiffness constant. In fact, the data in Fig.\ref{csivsT} can be nicely fit with $\rho_s= 0.79\pm
0.05\times 1.15 J_C/2\pi= 1.66\pm 0.1$ K.

Recently H\"artel et al. \cite{Hartel} have calculated the temperature dependence of $\xi(T)$ for $J_2\leq 0.44
|J_1|$ (i.e. $r\geq -0.44$), namely for the part of the phase diagram adjacent to the one experimentally
investigated here and in Ref.\onlinecite{PC2}. They found that $\xi(T)$ diverges exponentially with decreasing
temperature with an effective spin stiffness $\rho_s\simeq -(J_1 + 2J_2)/8$, which vanishes on approaching
$r\simeq -0.5$, namely the region with no long-range magnetic order.\cite{Hartel} SrZnVO(PO$_4$)$_2$, however, is
characterized by $r\simeq -1.15$ and it is not clear if the previous expression for the spin stiffness can still
be used. Nevertheless, if one considers that also for the compounds with $r\leq -0.5$ an analogous expression
$\rho_s\simeq +(J_1 + 2J_2)/8$ could hold, one would derive for SrZnVO(PO$_4$)$_2$ an effective spin stiffness
$\rho_s= 1.23\pm 0.16$ K, a value which is close to the one experimentally determined here (Fig.\ref{csivsT}).
Accordingly, in the absence of a theoretical calculation, one would be tempted to argue that on both sides of the
critical point around $r\simeq -0.5$ the correlation length diverges exponentially on cooling with an effective
spin stiffness $\rho_s\simeq |(J_1 + 2J_2)/8|$, progressively vanishing as $r\rightarrow 0.5$.

\begin{figure}[h!]
\vspace{7cm} \includegraphics{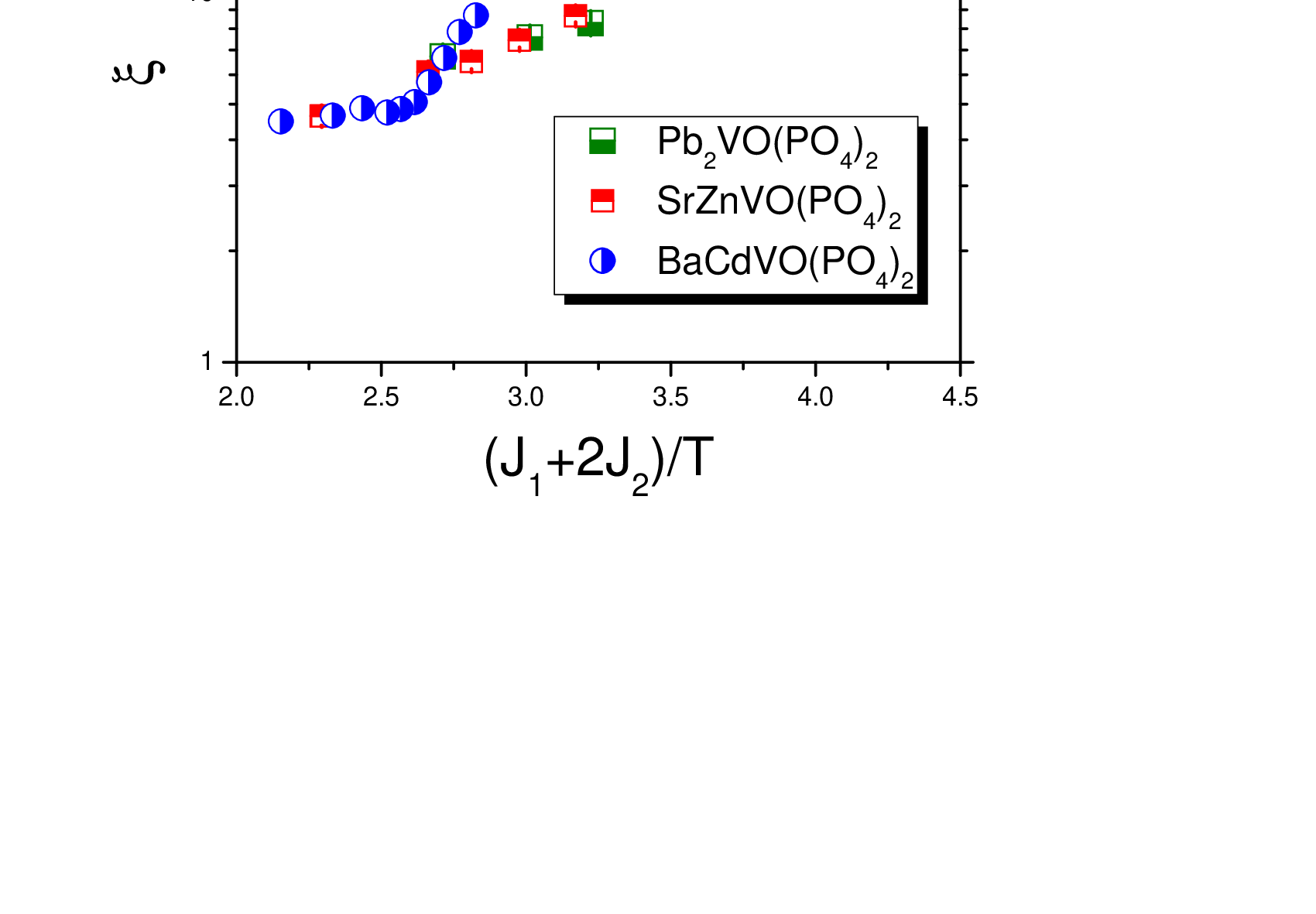}
\caption{\label{confrJ1J2} The in-plane correlation length derived from $\lambda(T)$ is reported as a function of
$(J_1+2J_2)/T$ for SrZnVO(PO$_4$)$_2$ ($(J_1+2J_2)= 9.73$ K), BaCdVO(PO$_4$)$_2$ ($(J_1+2J_2)= 2.8$
K)\cite{BaCd,Lucia} and Pb$_2$VO(PO$_4$)$_2$ ($(J_1+2J_2)= 13.7$ K).\cite{Pb2}}
\end{figure}

Now, it is rather interesting to compare the behaviour of $\xi$ in SrZnVO(PO$_4$)$_2$ with the one in
BaCdVO(PO$_4$)$_2$ ($r\simeq -0.9$ \cite{BaCd}) and in Pb$_2$VO(PO$_4$)$_2$ ($r\simeq -1.8$ \cite{Pb2}), derived
from the temperature dependence of $\lambda(T)$.\cite{PC2} Since in the latter two compounds it was not possible
to determine the absolute value of $\xi$ we assumed that $\lambda\sim \xi$ and rescaled the values of $\lambda$ so
that for $T\simeq J_C/2$, when $\xi\rightarrow 1$, $\xi$ is the same in all compounds. The corresponding data are
reported in Fig. \ref{xiJ2J1}. One notices that indeed $\xi$ decreases as $r\rightarrow -0.5$, however, it is also
noticed that while for SrZnVO(PO$_4$)$_2$ the correlation length diverges exponentially over a wide $T$ range ,
this is not the case for the other two systems. In fact, it has been pointed out that the behaviour of
Pb$_2$VO(PO$_4$)$_2$ is more characteristic of a 2D XY system,\cite{PC2} while in BaCdVO(PO$_4$)$_2$ possibly
nematic correlations appear, leading to a logarithmic increase of $\xi$ on cooling.\cite{PC2} Moreover, deviations
associated with the critical behaviour are observed on approaching $T_C$. Hence, it appears that although in
general the system becomes less correlated as $r\rightarrow -0.5$ the correct analytical form of the correlation
function is not simply exponential and that details taking into account the presence of a possible XY character or
of nematic correlations should be considered. Nevertheless, it is interesting to observe that in spite of the
functional form, the characteristic energy scale describing the growth of the in-plane correlation length appears
to scale as $J_1+2J_2$ far from $T_C$. In fact, if one now plots the $\xi$ data when $\xi\gg 1$ for the different
compounds as a function of $(J_1+2J_2)/T$, one observes a reasonable overlap between the data of the different
compounds until when the XY character or the inter-layer coupling do not give rise to a critical enhancement of
the correlations on approaching $T_C$. It is noticed that also in this plot the magnitude of $\xi$ in
Pb$_2$VO(PO$_4$)$_2$ and BaCdVO(PO$_4$)$_2$ has been rescaled in order to match the one quantitatively derived for
SrZnVO(PO$_4$)$_2$. Hence a definite answer on the validity of the scaling would require an independent
quantitative estimate of $\xi$ also for those two compounds.

\begin{figure}[h!]
\vspace{7cm} \includegraphics{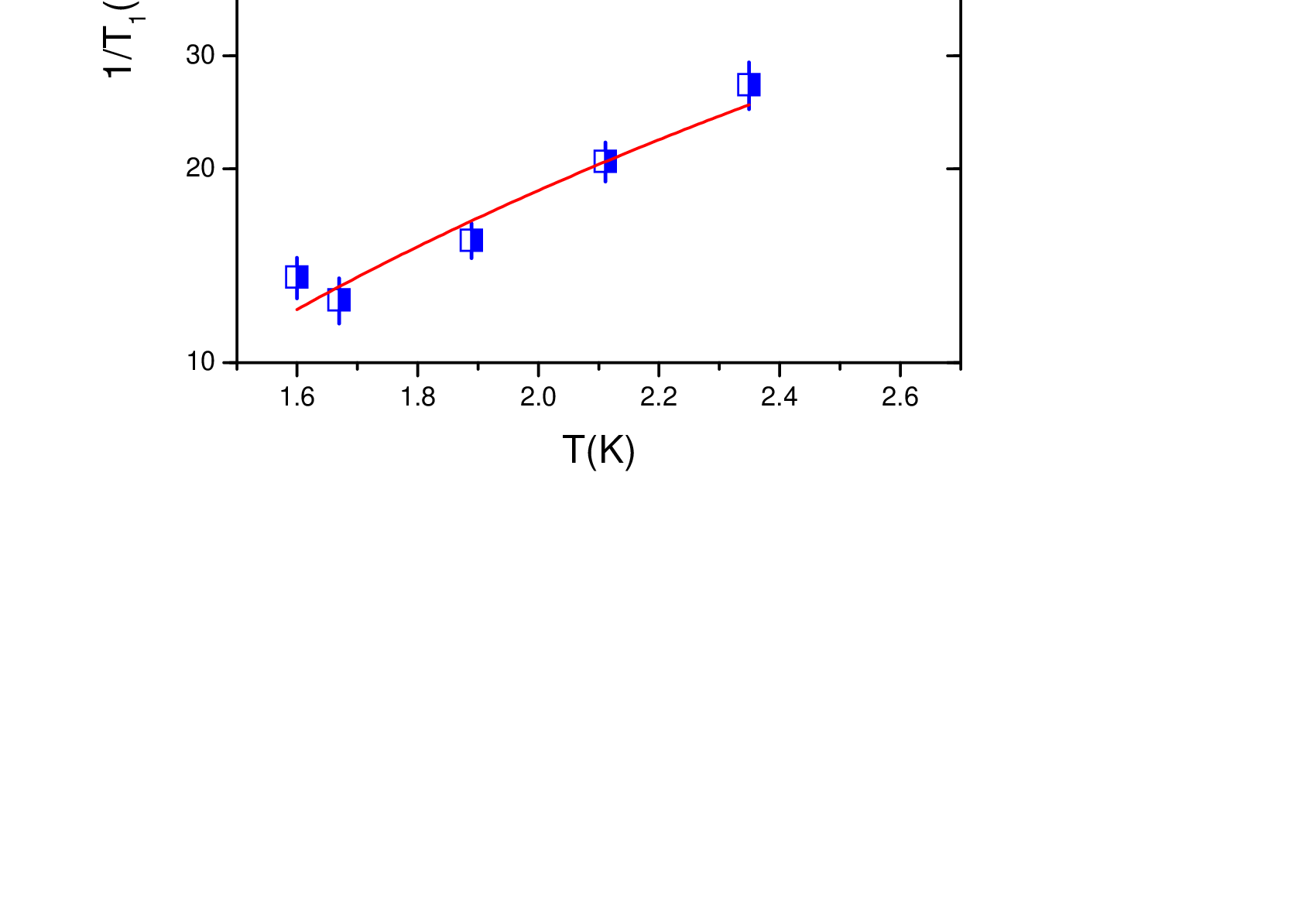} \caption{\label{T1magn} The
temperature dependence of $^{31}$P1 $1/T_1$ in SrZnVO(PO$_4$)$_2$ in the columnar ground-state is reported. The
solid line shows the best fit according to a power-law behaviour $1/T_1\sim T^b$ with $b= 1.9\pm 0.3$.}
\end{figure}

Finally, we shall discuss the behaviour of the spin-lattice relaxation rate in the magnetically ordered columnar
phase. Below $T_C$ one observes a marked decrease of $1/T_1$ which in the low-temperature limit should be ascribed
to the vanishing of the two-magnon Raman relaxation processes.\cite{Pincus} If the gap in the magnon dispersion
curve is negligible one would expect a power-law behaviour of $1/T_1$ with a power law exponent depending on the
magnetic lattice dimensionality and on the analytical form of the magnon dispersion curve.\cite{Pincus} In case of
a linear dispersion curve, neglecting the presence of a gap in the spin-wave dispersion, for a quasi-2D system one
would expect $1/T_1\sim T^2$. Here we find that $1/T_1\sim T^b$ with $b= 1.9\pm 0.3$, in reasonable agreement with
the theoretical expectations and with the behaviour of the sublattice magnetization derived from $\mu^+$SR
measurements.

\section{Conclusions}

In conclusion we have determined quantitatively the temperature dependence of the in-plane correlation length
$\xi$ in SrZnVO(PO$_4$)$_2$, a frustrated $S=1/2$ magnet on a square lattice with $r\simeq -1.15$, by means of
nuclear and muon spin-lattice relaxation rate measurements. It has been shown that $\xi$ diverges exponentially on
cooling with a reduced spin stiffness which appears to roughly scale as $|J_1+ 2J_2|$. A comparison with the
results previously obtained by our group on other systems with $r< 0$ appears to support this scaling of the spin
stiffness even if an accurate description of $\xi$ on approaching the transition to the columnar ground-state
should take into account the spin anisotropy and interlayer couplings.

\section*{Acknowledgements}

The technical assistance by Sean Giblin during the measurements at ISIS is gratefully acknowledged. The research
activity in Pavia was supported by Fondazione Cariplo (Grant N. 2008-2229) research funds.




\begin{references}
\bibitem{Lucia} Susceptibility and magnetization vs. field measurements in our BaCdVO(PO$_4$)$_2$
sample confirmed, within the error bars, the values reported in Ref.\onlinecite{BaCd}.
\end{references}
\end{document}